\documentclass[10pt,conference]{IEEEtran}

\usepackage{epsfig,rotating,setspace,latexsym,amsmath,epsf,amssymb,amsfonts,bm,theorem,subfigure,epstopdf}
\DeclareMathOperator*{\argminA}{arg\,min}
\usepackage{cite}
\usepackage{bbm}
\usepackage[ruled,vlined]{algorithm2e}
\usepackage{color}

\usepackage[T1]{fontenc}

\IEEEoverridecommandlockouts
\allowdisplaybreaks

\begin{document}
\bstctlcite{IEEEexample:BSTcontrol}

	%
	\title{Adversarial Attacks with Multiple Antennas Against Deep Learning-Based Modulation Classifiers}

		\author{\IEEEauthorblockN{Brian Kim, Yalin E. Sagduyu, Tugba Erpek, Kemal Davaslioglu, and Sennur Ulukus}
		\thanks{Brian Kim and Sennur Ulukus are with University of Maryland, College Park, MD, USA; Email: \{bkim628, ulukus\}@umd.edu.}
		\thanks{Yalin E. Sagduyu and Kemal Davaslioglu are with Intelligent Automation, Inc., Rockville, MD, USA; Email: \{ysagduyu, kdavaslioglu\}@i-a-i.com.} \thanks{Tugba Erpek is with Virginia Tech., Hume Center, Arlington, VA, USA, and Intelligent Automation, Inc., Rockville, MD, USA; Email: {terpek}@vt.edu.
		}
		\thanks{This effort is supported by the U.S. Army Research Office under contract W911NF-17-C-0090. The content of the information does not necessarily reflect the position or the policy of the U.S. Government, and no official endorsement should be inferred.} 
	}

	
\maketitle

\begin{abstract}
We consider a wireless communication system, where a transmitter sends signals to a receiver with different modulation types while the receiver classifies the modulation types of the received signals using its deep learning-based classifier. Concurrently, an adversary transmits adversarial perturbations using its multiple antennas to fool the classifier into misclassifying the received signals. From the adversarial machine learning perspective, we show how to utilize multiple antennas at the adversary to improve the adversarial (evasion) attack performance. Two main points are considered while exploiting the multiple antennas at the adversary, namely the power allocation among antennas and the utilization of channel diversity. First, we show that multiple independent adversaries, each with a single antenna cannot improve the attack performance compared to a single adversary with multiple antennas using the same total power. Then, we consider various ways to allocate power among multiple antennas at a single adversary such as allocating power to only one antenna, and proportional or inversely proportional to the channel gain. By utilizing channel diversity, we introduce an attack to transmit the adversarial perturbation through the channel with the largest channel gain at the symbol level. We show that this attack reduces the classifier accuracy significantly compared to other attacks under different channel conditions in terms of channel variance and channel correlation across antennas. Also, we show that the attack success improves significantly as the number of antennas increases at the adversary that can better utilize channel diversity to craft adversarial attacks.
\end{abstract}

\section{Introduction}
 
Recent advances in deep learning (DL) have enabled numerous applications in different domains such as computer vision \cite{vision1} and speech recognition \cite{Goodfellow1}. Upon the success of these applications, DL has been also applied to wireless communications where the high-dimensional spectrum data is analyzed by deep neural networks (DNNs) while accounting for unique characteristics of the wireless medium such as waveform, channel, interference, and traffic effects \cite{Oshea2,Oshea3,erpek1}. Examples of wireless communication applications that benefit from DL include waveform design \cite{erpek1}, spectrum sensing \cite{Davaslioglu2}, and signal classification \cite{Oshea2}.
 
Despite the benefits of DL, DNNs are known to be susceptible to adversarial manipulation of their input causing incorrect outputs such as classification labels as demonstrated first in computer vision applications \cite{Szegedy1}. Therefore, machine learning in the presence of adversaries has received significant attention in the computer vision domain and has been extensively studied in the context of adversarial machine learning \cite{Vorobeychik1}. 
Different types of attacks built upon adversarial machine learning are feasible in wireless communication systems such as exploratory attacks \cite{Terpek}, adversarial attacks \cite{Larsson2}, poisoning attacks \cite{Sagduyu1}, membership inference attacks \cite{MIA}, and Trojan attacks \cite{Davaslioglu1}. These attacks have the advantage of being stealthier than conventional jamming attacks that typically  add interference directly to data transmissions without specifically targeting the underlying machine learning applications \cite{Sagduyu2008}. 

In this paper, we focus on adversarial attacks (also known as evasion attacks) which correspond to adding small perturbations to the original input of the DNNs in order to cause misclassification. These perturbations are not just random but are carefully crafted to fool the DNNs. 
Adversarial attacks on modulation classifier \cite{Oshea2} of wireless signals have been studied in \cite{Larsson2} where fast gradient method (FGM) \cite{Kurakin1} is used to create adversarial perturbations.
In \cite{Kokalj2,Kokalj3, Flowers1}, it has been shown that the modulation classifier is vulnerable to various forms of adversarial attacks in the AWGN channel. Adversarial attacks in the presence of realistic channel effects and broadcast transmissions have been studied in \cite{Kim1,Kim2}. The attack setting has been also extended to incorporate communication error performance \cite{Gunduz1} and covertness \cite{Kim5G}. 

Our goal in this paper is to investigate the use of multiple antennas to generate multiple concurrent perturbations over different channel effects (subject to a total power budget) to the input of a DNN-based modulation classifier at a wireless receiver. This problem setting is different from computer vision applications of adversarial attacks that are limited to a single perturbation that can be directly added to the DNN's input without facing uncertainties such as channel effects. We assume that the adversary has multiple antennas to transmit adversarial perturbations in the presence of realistic channel effects and aims to decrease the accuracy of a modulation classifier. As shown in \cite{Kim1}, transmitting random (e.g., Gaussian) noise to decrease the accuracy of the classifier at the receiver is ineffective as an adversarial attack, since random noise cannot manipulate the input to the DNN in a specific direction as needed in an adversarial attack. Therefore, increasing the perturbation power with random noise transmitted over multiple antennas remains ineffective. Instead, the adversary needs to carefully craft the adversarial perturbation for each antenna. 

We design a white-box attack where the adversary knows the receiver's classifier architecture, input at the receiver, and the channel between the adversary and the receiver. The adversary signal is time-aligned with the transmitted signal and uses the maximum received perturbation power (MRPP) attack that was introduced in \cite{Kim1}. First, we show that just increasing the number of individual adversaries with single antennas (located at different positions) does not improve the attack performance. Next, we consider the use of multiple antennas at a single adversary and propose different methods to allocate power among antennas at the adversary and to exploit the channel diversity. We first propose a genie-aided adversarial attack where the adversary selects one antenna to transmit the perturbation such that it would result in the worst classification performance depending on the channel condition over the entire symbol block (that corresponds to the input to the DNN at the receiver). Then, we consider transmitting with all the antennas at the adversary where the power allocation is based on the channel gains, either proportional or inversely proportional to the channel gains. However, these attacks remain ineffective. We propose the elementwise maximum channel gain (EMCG) attack to utilize the channel diversity more efficiently by selecting the antenna with the best channel gain at the symbol level to transmit perturbations. 



We show that the EMCG attack outperforms other attacks and effectively uses channel diversity provided by multiple antennas to cause misclassification at the receiver. This attack improvement remains effective regardless of the channel variance or correlation between channels, whereas the proportional to the channel gain (PCG) attack is greatly affected by the correlation between channels. Finally, we show that increasing the number of antennas at the adversary significantly improves the attack performance by better exploiting the channel diversity to craft and transmit adversarial perturbations.  

The rest of the paper is organized as follows. Section \ref{sec:SystemModel} provides the system model. Section \ref{sec:section3} introduces adversarial attacks using multiple antennas. Section \ref{sec:sim} presents simulation results. Section \ref{sec:Conclusion} concludes the paper. 

\section{System Model} \label{sec:SystemModel}

\begin{figure}[t]
	\centerline{\includegraphics[width=0.75\linewidth]{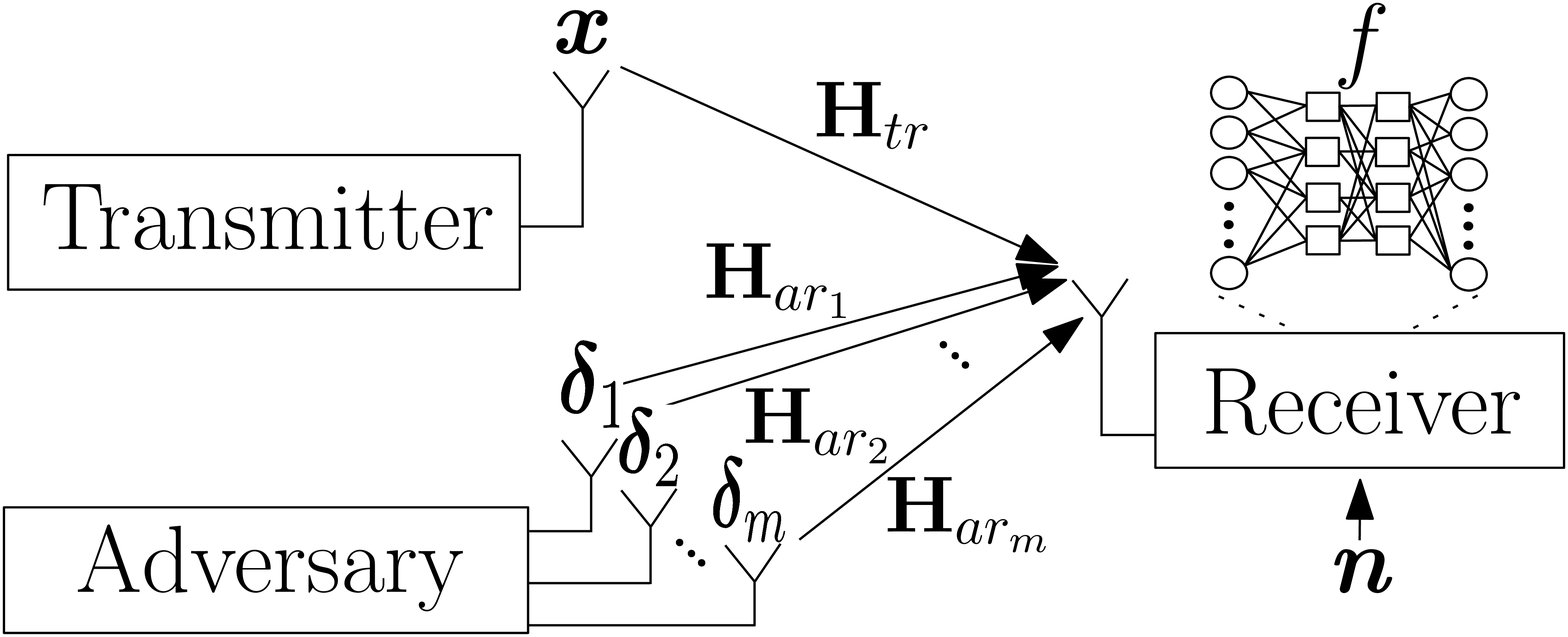}}
	\caption{System model.}
	\label{sys}
\end{figure}

We consider a wireless communication system that consists of a transmitter, a receiver, and an adversary as shown in Fig. \ref{sys}. Both the transmitter and the receiver are equipped with a single antenna. The receiver uses a pre-trained DL-based classifier on the received signals to classify the modulation type that is used at the transmitter. The adversary has $m$ antennas to launch a white-box adversarial attack to cause misclassification at the receiver. The white-box attack can be considered as an upper-bound for other attacks with limited information. The assumptions on the knowledge of the adversary can be relaxed as shown in \cite{Kim1}.

The DNN classifier at the receiver is denoted by $f(\cdot;\boldsymbol{\theta}): \mathcal{X} \rightarrow \mathbb{R}^{C}$, where $\boldsymbol{\theta}$ is the set of parameters of the DNN decided in the training phase and $C$ is the number of modulation types. Note $\mathcal{X} \subset \mathbb{C}^{p}$, where $p$ is the dimension of the complex-valued I/Q (in-phase/quadrature) inputs to the DNN that can also be represented by concatenation of two real-valued inputs. A modulation type $\hat{l}(\boldsymbol{x},\boldsymbol{\theta}) = \arg \max_{k} f_{k}(\boldsymbol{x},\boldsymbol{\theta})$ is assigned by $f$ to input $\boldsymbol{x}\in \mathcal{X}$. In this formulation, $f_{k}(\boldsymbol{x},\boldsymbol{\theta})$ is the output of classifier $f$ corresponding to the $k$th modulation type.

The channel from the transmitter to the receiver is $\boldsymbol{h}_{tr}$, and the channel from the $i$th antenna of the adversary to the receiver is  $\boldsymbol{h}_{ar_{i}}$, where $\boldsymbol{h}_{tr} = [h_{tr,1}, h_{tr,2},\cdot \cdot \cdot,h_{tr,p}]^{T}\in \mathbb{C}^{p\times 1}$ and $\boldsymbol{h}_{ar_{i}} = [h_{ar_{i},1}, h_{ar_{i},2},\cdot \cdot \cdot,h_{ar_{i},p}]^{T}\in \mathbb{C}^{p\times 1}$. If the transmitter transmits $\boldsymbol{x}$, the receiver receives $\boldsymbol{r}_{t} = \mathbf{H}_{tr} \boldsymbol{x}+\boldsymbol{n}$, if there is no adversarial attack, or receives $\boldsymbol{r}_{a} = \mathbf{H}_{tr} \boldsymbol{x}+ \sum_{i=1}^{m}\mathbf{H}_{ar_{i}}\boldsymbol{\delta}_{i}+\boldsymbol{n}$, if the adversary transmits the perturbation signal $\boldsymbol{\delta}_{i}$ at the $i$th antenna, where $\mathbf{H}_{tr} = \mbox{diag} \{h_{tr,1},\cdot\cdot\cdot, h_{tr,p}\}\in \mathbb{C}^{p\times p}, \mathbf{H}_{ar_{i}} = \mbox{diag}\{h_{ar_{i},1},\cdot \cdot\cdot,h_{ar_{i},p}\}\in \mathbb{C}^{p \times p}$, $\boldsymbol{\delta}_{i}\in \mathbb{C}^{p\times1}$ and $\boldsymbol{n}\in \mathbb{C}^{p\times 1}$ is complex Gaussian noise. For a stealth attack, the adversarial perturbations on antennas are constrained as $\sum_{i=1}^{m}\|\boldsymbol{\delta}_{i}\|^{2}_{2}\le P_{\textit{max}}$ for some suitable power $P_{\textit{max}}$. To determine these perturbations with respect to the transmitted signal $\boldsymbol{x}$ from the transmitter, the adversary solves the following optimization problem.
\begin{align}\label{optimization_eq}
\argminA_{\{\boldsymbol{\delta}_{i}\}} & \quad \; \sum_{i=1}^{m}\|\boldsymbol{\delta}_{i}\|^{2}_{2} \nonumber\\
\mbox{subject to} & \quad \hat{l}(\boldsymbol{r}_{tr},\boldsymbol{\theta}) \ne \hat{l}(\boldsymbol{r}_{ar_{i}}(\boldsymbol{\delta}_{i}),\boldsymbol{\theta}), \quad i = 1,2, ...,m \nonumber \\
& \quad \sum_{i=1}^{m}\|\boldsymbol{\delta}_{i}\|^{2}_{2} \le P_{\textit{max}}.
\end{align}

In (\ref{optimization_eq}), the objective is to minimize the perturbation power subject to two constraints where the receiver misclassifies the received signal and the budget for perturbation power is not exceeded. However, solving optimization problem (\ref{optimization_eq}) is difficult because of the inherent structure of the DNN. Thus, different methods have been proposed to approximate the adversarial perturbation. For instance, FGM is a computationally efficient method for generating adversarial attacks by linearizing the loss function of the DNN classifier. We denote the loss function of the model by $L(\boldsymbol{\theta},\boldsymbol{x},\boldsymbol{y})$, where $\boldsymbol{y}\in \{0,1 \}^C$ is the one-hot encoded class vector. Then, FGM linearizes this loss function in a neighborhood of $\boldsymbol{x}$ and uses this linearized function for optimization. Since the adversary uses more than one antenna, the adversary needs to utilize the diversity of channels to craft more effective perturbations. For that purpose, we introduce different approaches in Section \ref{sec:section3}.


\section{Adversarial Attacks using Multiple Antennas} \label{sec:section3}
In this section, we introduce different methods to utilize multiple antennas at the adversary to improve the attack performance. Note that the adversary can allocate power differently to each antenna and increase the channel diversity by using multiple antennas. Throughout the paper, we apply the targeted MRPP attack in \cite{Kim1} to generate an attack at the adversary. The MRPP attack searches over all modulation types to cause misclassification at the receiver and chooses one modulation type that needs the least power to cause the misclassification.  

\begin{algorithm}[t]
	\DontPrintSemicolon
	\SetAlgoLined
	\label{prop_com}
	Inputs: input $\boldsymbol{r}_{tr}$, desired accuracy $\varepsilon_{acc}$, power constraint $P_{\textit{max}}$ and model of the classifier $L(\boldsymbol{\theta},\cdot,\cdot)$\\
	Initialize: $ \boldsymbol{\varepsilon}\leftarrow \boldsymbol{0}^{C \times 1}$,
	 	$w_{i} = \frac{\|\mathbf{h}_{ar_{i}}\|_{2}}{\sum_{j=1}^{m}\|\mathbf{h}_{ar_{j}}\|_{2}}, i = 1, \cdots, m$ \\
	
	\For{class-index $c$ in range($C$)}{
		$\varepsilon_{\textit{max}} \leftarrow \sqrt{P_{\textit{max}}}, \varepsilon_{min} \leftarrow 0$\\
			\For{$i=1\;\emph{\KwTo} \; m$ }{
			{$\boldsymbol{\delta}_{i}^{c} =\frac{\mathbf{H}^{*}_{ar_{i}}\nabla_{\boldsymbol{x}}L(\boldsymbol{\theta},\boldsymbol{r}_{tr},\boldsymbol{y}^{c})}{(\|\mathbf{H}^{*}_{ar_{i}}\nabla_{\boldsymbol{x}}L(\boldsymbol{\theta},\boldsymbol{r}_{tr},\boldsymbol{y}^{c})\|_{2})}$\\}
			
		}
		
		\While{$\varepsilon_{\textit{max}}-\varepsilon_{min} > \varepsilon_{acc}$}{
			$\varepsilon_{avg} \leftarrow (\varepsilon_{\textit{max}}+\varepsilon_{min})/2$\\
			$\boldsymbol{x}_{adv} \leftarrow \boldsymbol{x} - \varepsilon_{avg}\sum_{i=1}^{m}w_{i}\mathbf{H}_{ar_{i}}\boldsymbol{\delta}_{i}^{c}$\\
			\lIf{$\hat{l}(\boldsymbol{x}_{adv})== l_{\textit{true}}$}{$\varepsilon_{min}\leftarrow \varepsilon_{avg}$}
			\lElse{$\varepsilon_{\textit{max}}\leftarrow \varepsilon_{avg}$}
		}
		$\boldsymbol{\varepsilon}[c] = \varepsilon_{\textit{max}}$\\
}
	
		$\textit{target} = \arg\min \boldsymbol{\varepsilon}$, 
		$\boldsymbol{\delta}_{i} = \boldsymbol{\varepsilon}[\textit{target}]w_i\boldsymbol{\delta}_{i}^{\textit{target}}$ for $\forall i$

	\caption{PCG attack with common target}
\end{algorithm}

\subsection{Single-Antenna Genie-Aided (SAGA) Attack}
We first begin with an attack where the adversary allocates all the power to only one antenna for the entire symbol block of an input to the classifier at the receiver as shown in Fig. \ref{illus}(a). In this attack, we assume that the adversary is aided by a Genie and thus knows in advance the best antenna out of $m$ antennas that causes a misclassification. Then, the Genie-aided adversary puts all the power to that one specific antenna to transmit adversarial perturbation.

\subsection{Proportional to Channel Gain (PCG) Attack}
To exploit the channel with the better channel gain, the adversary allocates more power to better channels. Specifically, the power allocation for the $i$th antenna is proportional to the channel gain $\|\mathbf{h}_{ar_{i}}\|_{2}$. The adversarial perturbation that is transmitted by each antenna is generated using the MRPP attack as before and transmitted with the power allocated to each antenna. During the attack generation process, the adversary can set the common target modulation type of misclassification (each perturbation aims to cause misclassification of signals to a common target label) or independent target (each perturbation aims to cause misclassification of signals to its own target label) for each antenna.

\subsubsection{PCG attack with common target}
The adversary sets a common target modulation type for all antennas to cause specific misclassification at the receiver. The adversary determines the common target modulation type which needs the least power to fool the receiver. The details are presented in Algorithm \ref{prop_com}. 

\subsubsection{PCG attack with independent targets}
For the $i$th antenna, the adversary decides the individual target modulation type for perturbation $\boldsymbol{\delta}_{i}$. Each antenna independently chooses the target modulation type which uses the least power to cause misclassification at the receiver. These modulation types may differ from each other. By setting individual target modulation type for each antenna, the adversary can exploit the channel since each antenna chooses what is best for itself. The details are presented in Algorithm \ref{prop_ind}.

\begin{algorithm}[t]
	\DontPrintSemicolon
	\SetAlgoLined
	\label{prop_ind}
	Inputs: input $\boldsymbol{r}_{tr}$, desired accuracy $\varepsilon_{acc}$, power constraint $P_{\textit{max}}$ and model of the classifier $L(\boldsymbol{\theta},\cdot,\cdot)$\\
	Initialize: $ \boldsymbol{\varepsilon}\leftarrow \boldsymbol{0}^{C \times 1}$, $w_{i} = \frac{\|\mathbf{h}_{ar_{i}}\|_{2}}{\sum_{j=1}^{m}\|\mathbf{h}_{ar_{j}}\|_{2}}, i = 1, \cdots, m$ \\
	\For{$i=1\;\emph{\KwTo} \; m$ }{
		\For{class-index $c$ in range($C$)}{
			$\varepsilon_{\textit{max}} \leftarrow \sqrt{P_{\textit{max}}}, \varepsilon_{min} \leftarrow 0$\\
			
			{$\boldsymbol{\delta}_{i}^{c} =\frac{\mathbf{H}^{*}_{ar_{i}}\nabla_{\boldsymbol{x}}L(\boldsymbol{\theta},\boldsymbol{r}_{tr},\boldsymbol{y}^{c})}{(\|\mathbf{H}^{*}_{ar_{i}}\nabla_{\boldsymbol{x}}L(\boldsymbol{\theta},\boldsymbol{r}_{tr},\boldsymbol{y}^{c})\|_{2})}$\\}

			\While{$\varepsilon_{\textit{max}}-\varepsilon_{min} > \varepsilon_{acc}$}{
				$\varepsilon_{avg} \leftarrow (\varepsilon_{\textit{max}}+\varepsilon_{min})/2$\\
				$\boldsymbol{x}_{adv} \leftarrow \boldsymbol{x} - \varepsilon_{avg}\sum_{i=1}^{m}w_{i}\mathbf{H}_{ar_{i}}\boldsymbol{\delta}_{i}^{c}$\\
				\lIf{$\hat{l}(\boldsymbol{x}_{adv})== l_{\textit{true}}$}{$\varepsilon_{min}\leftarrow \varepsilon_{avg}$}
				\lElse{$\varepsilon_{\textit{max}}\leftarrow \varepsilon_{avg}$}
			}
		}
		$\boldsymbol{\varepsilon}[c] = \varepsilon_{\textit{max}}$\\
		$\textit{target} = \arg\min \boldsymbol{\varepsilon}$, 
		$\boldsymbol{\delta}_{i} = \boldsymbol{\varepsilon}[\textit{target}]w_i\boldsymbol{\delta}_{i}^{\textit{target}}$
	}
	
	\caption{PCG attack with independent targets}
\end{algorithm}

\subsection{Inversely Proportional to Channel Gain (IPCG) Attack}
In contrast to the PCG attack, the adversary allocates more power to weak channels to compensate for the loss over the weak channels, i.e., inversely proportional to the channel gain. The perturbations that are transmitted by each antenna are generated using the MRPP attack and the power for each antenna is determined to be inversely proportional to the channel gain. As in the PCG attack, the IPCG attack can be also crafted with common target or independent targets for all antennas. The algorithm is the same as Algorithm \ref{prop_com} for common target and Algorithm \ref{prop_ind} for the independent targets except that $w_{i}$ changes to be inversely proportional to the channel, i.e., $w_{i} = \frac{1}{\|\mathbf{h}_{ar_{i}}\|_{2} \left(\frac{1}{\sum_{j=1}^{m}\|\mathbf{h}_{ar_{j}}\|_{2}}\right)}, i = 1, \cdots, m$.
%
%

\begin{algorithm}[t]
	\DontPrintSemicolon
	\SetAlgoLined
	\label{alg2}
	Inputs: input $\boldsymbol{r}_{tr}$, desired accuracy $\varepsilon_{acc}$, power constraint $P_{\textit{max}}$ and model of the classifier $L(\boldsymbol{\theta},\cdot,\cdot)$\\
	Initialize: $ \boldsymbol{\varepsilon}\leftarrow \boldsymbol{0}^{C \times 1}$, $k \leftarrow \boldsymbol{0}^{p \times 1}$, $\boldsymbol{\delta}_{i} \leftarrow \boldsymbol{0}^{p \times 1}$ for $\forall i$ \\
	\For{$i=1\;\emph{\KwTo} \; p$}{
		$h_{vir,i} = \max\{ \|h_{ar_{1},i}\|_{2},\cdots, \|h_{ar_{m},i}\|_{2}\}$\\
		$k[i] = \arg\max\{ \|h_{ar_{1},i}\|_{2}, \cdots, \|h_{ar_{m},i}\|_{2}\}$
	}
	{Virtual channel : $\mathbf{H}_{vir} = \mbox{diag}\{h_{vir,1},\cdots,h_{vir,p}\}$\\}
	\For{class-index $c$ in range($C$)}{
		$\varepsilon_{\textit{max}} \leftarrow \sqrt{P_{\textit{max}}}, \varepsilon_{min} \leftarrow 0$\\
		$\boldsymbol{\delta}^{c} =\frac{\mathbf{H}^{*}_{vir}\nabla_{\boldsymbol{x}}L(\boldsymbol{\theta},\boldsymbol{r}_{tr},\boldsymbol{y}^{c})}{(\|\mathbf{H}^{*}_{vir}\nabla_{\boldsymbol{x}}L(\boldsymbol{\theta},\boldsymbol{r}_{tr},\boldsymbol{y}^{c})\|_{2})}$\\
		\While{$\varepsilon_{\textit{max}}-\varepsilon_{min} > \varepsilon_{acc}$}{
			$\varepsilon_{avg} \leftarrow (\varepsilon_{\textit{max}}+\varepsilon_{min})/2$\\
			$\boldsymbol{x}_{adv} \leftarrow \boldsymbol{x} - \varepsilon_{avg}\mathbf{H}_{vir}\boldsymbol{\delta}^{c}$\\
			\lIf{$\hat{l}(\boldsymbol{x}_{adv})== l_{\textit{true}}$}{$\varepsilon_{min}\leftarrow \varepsilon_{avg}$}
			\lElse{$\varepsilon_{\textit{max}}\leftarrow \varepsilon_{avg}$}
		}
		$\boldsymbol{\varepsilon}[c] = \varepsilon_{\textit{max}}$\\
	}
	$\textit{target} = \arg\min \boldsymbol{\varepsilon}, \boldsymbol{\delta}^{vir} = \boldsymbol{\varepsilon}[\textit{target}]\boldsymbol{\delta}^{\textit{target}} $\\
	
	\For{$i=1\;\emph{\KwTo} \; p$}{$\boldsymbol{\delta}_{k[i]} =\boldsymbol{\delta}^{vir}[i]  $\\}
	Transmit $\boldsymbol{\delta}_{i}$, $i = 1,\cdots,m$
	\caption{EMCG attack}
\end{algorithm}

\begin{figure}[t]
	\centering
	\subfigure[]{\includegraphics[width=0.7\linewidth]{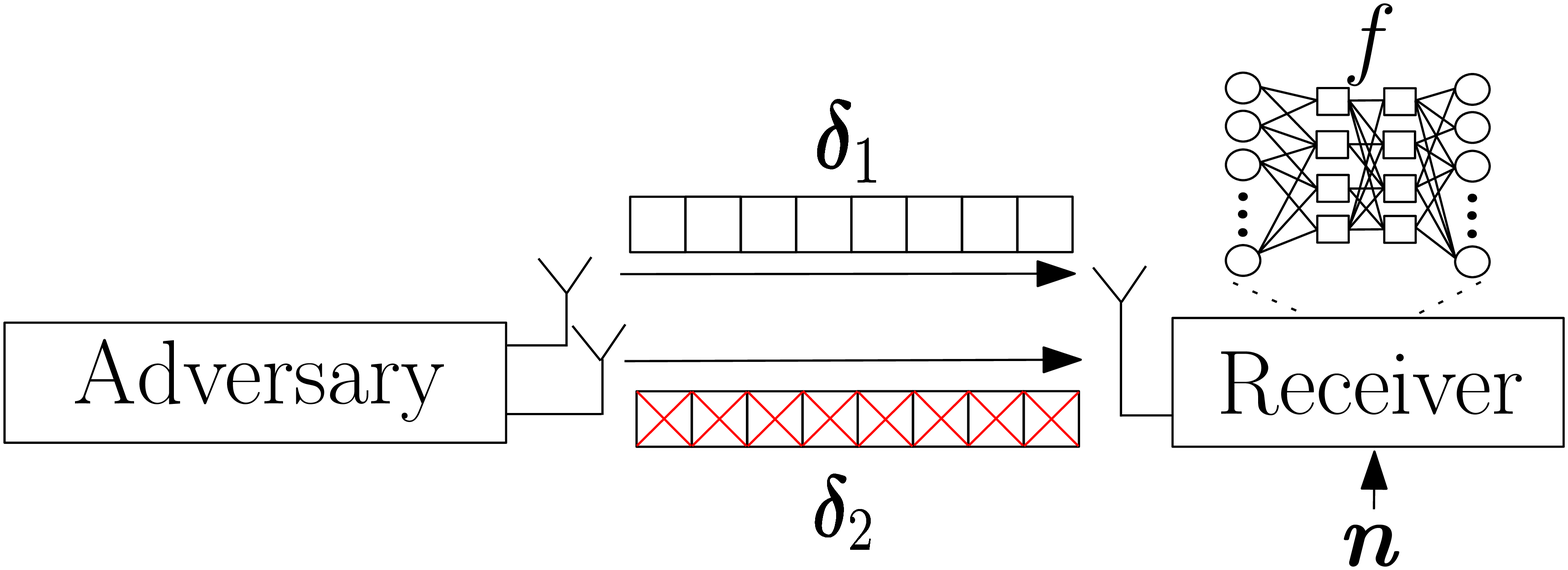} } 
	\subfigure[]{\includegraphics[width=0.7\linewidth]{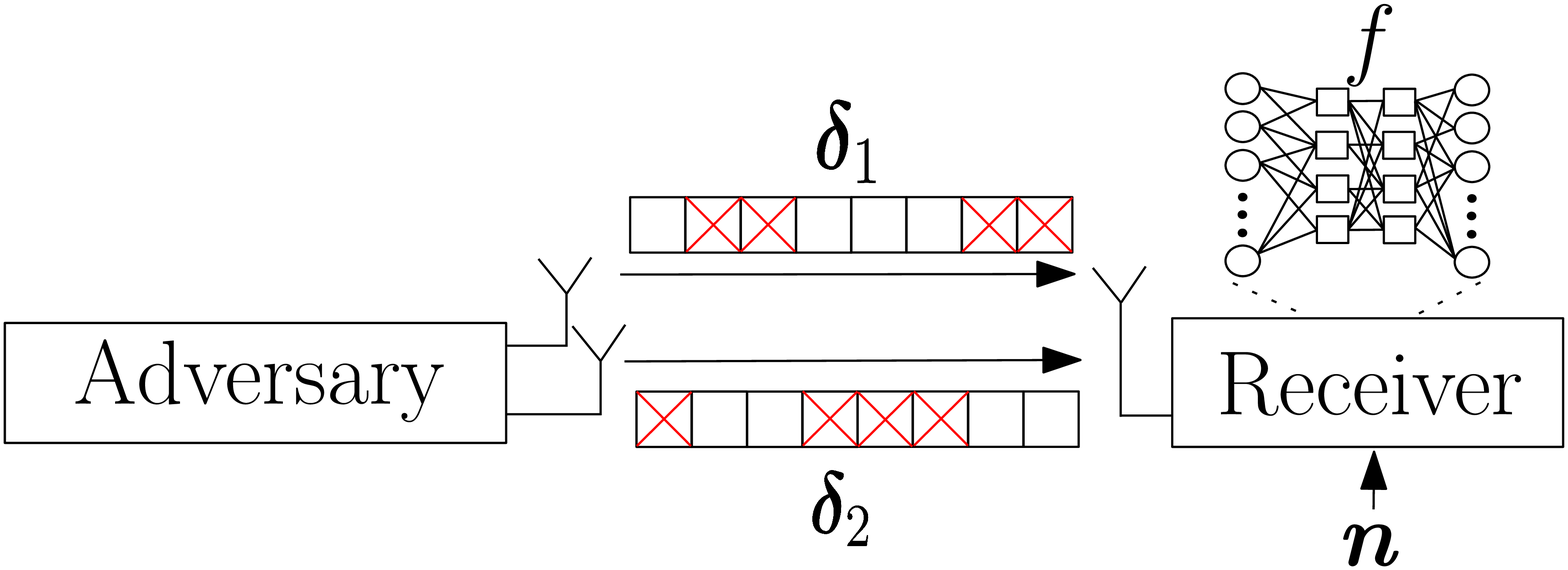} } 		
	\caption{Illustration of (a) SAGA attack and (b) EMCG attack. }
	\label{illus}
\end{figure}

\subsection{Elementwise Maximum Channel Gain (EMCG) Attack}
Unlike the previous attacks that considered the channel gain of the channel vector with dimension $p \times 1$ as a way to allocate power among antennas, the adversary in the EMCG attack considers the channel gain of each element of the channel to fully utilize the channel diversity as shown in Fig. \ref{illus}(b). First, the adversary compares the channel gain elementwise and selects one antenna that has the largest channel gain at each instance. Specifically, the adversary finds and transmits with the antenna $j^* = \arg\max_{j=1, \cdots m}\{ \|h_{ar_{j},t}\|_{2}\}$ that has the largest channel gain at instance $t$. Further, a virtual channel ${h}_{vir,t}$ at instance $t$ is defined as the channel with the largest channel gain among antennas. Then, the adversary generates the perturbation  $\boldsymbol{\delta}^{vir}$ with respect to $\boldsymbol{h}_{vir} = [h_{vir,1},\cdots,h_{vir,p}]^{T}$ using the MRPP attack and transmits each element of $\boldsymbol{\delta}^{vir}$ with the antenna that has been selected previously. The details are presented in Algorithm \ref{alg2}.


\section{Simulation Results} \label{sec:sim}

In this section, we compare the performances of the attacks introduced in Section \ref{sec:section3} (along with the MRPP attack from \cite{Kim1} where the adversary has a single antenna) to investigate how the number of antennas at the adversary affects the attack performance. Also, multiple adversaries that are each equipped with a single antenna and located at different positions are considered to motivate the need to craft attacks for the adversary with multiple antennas. 

\begin{figure}[t]
	\centerline{\includegraphics[width=0.8\linewidth]{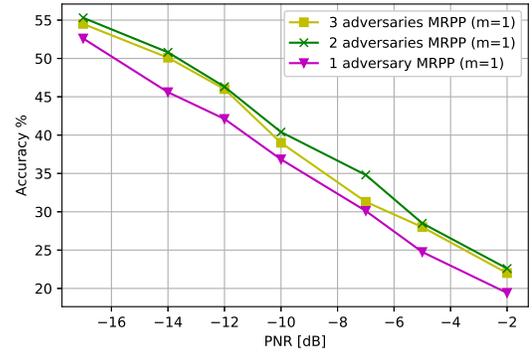}}
	\caption{Classifier accuracy with respect to the number of adversaries with single antenna.}
	\label{adv}
\end{figure}
To evaluate the performance, we use the VT-CNN2 classifier from \cite{Oshea1} as the modulation classifier (also used in \cite{Larsson2,Kim1}) where the classifier consists of two convolution layers and two fully connected layers, and train it with GNU radio ML dataset RML2016.10a \cite{dataset1}. The dataset contains 220,000 samples where half of the samples are used for training and the other half are used for testing. Each sample corresponds to one specific modulation type at a specific signal-to-noise ratio (SNR). There are 11 modulations which are BPSK, QPSK, 8PSK, QAM16, QAM64, CPFSK, GFSK, PAM4, WBFM, AM-SSB and AM-DSB. We follow the same setup of \cite{Oshea1}, using Keras with TensorFlow backend, where the input sample to the modulation classifier is 128 I/Q channel symbols.

\begin{figure}[t]
	\centerline{\includegraphics[width=1\linewidth]{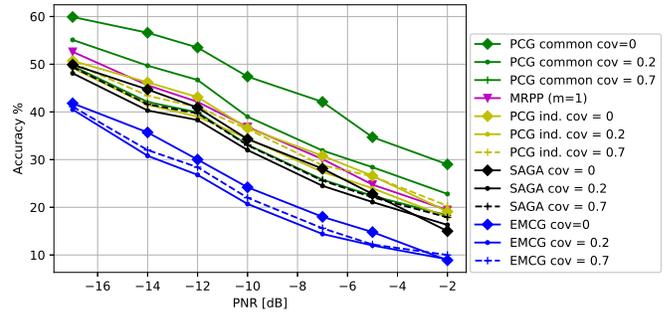}}
	\caption{Classifier accuracy with respect to different covariances of channels between antennas.}
	\label{cor}
\end{figure}

In the simulations, we introduce the channel between the $i$th antenna at the adversary and the receiver as a Rayleigh fading channel with path-loss and shadowing, i.e., $h_{ar_{i},j}=K(\frac{d_{0}}{d})^{\gamma}\psi h_{i,j}$ where $ K = 1, d_{0}=1, d=10, \gamma = 2.7, \psi \sim\ $Lognormal$(0,8)$ and ${h}_{i,j} \sim \mbox{Rayleigh}(0,1)$. We assume that channels between antennas are independent (except for Fig. \ref{cor}) and fix SNR as 10dB. We evaluate the attack performance as a function of the perturbation-to-noise ratio (PNR) from \cite{Larsson2}. The PNR represents the relative perturbation power with respect to the noise power. As the PNR increases, the power of the perturbation relatively increases compared to the noise power making the perturbation more likely to be detected by the receiver since it becomes more distinguishable from noise.

\begin{figure}[t]
	\centerline{\includegraphics[width=0.8\linewidth]{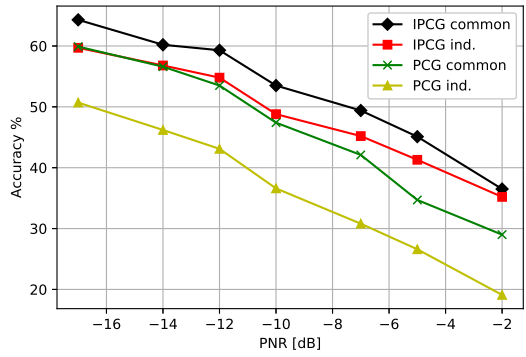}}
	\caption{Classifier accuracy when adversarial attacks with common target and independent targets are transmitted at the adversary.}
	\label{target}
\end{figure}

\begin{figure}[t]
	\centerline{\includegraphics[width=0.8\linewidth]{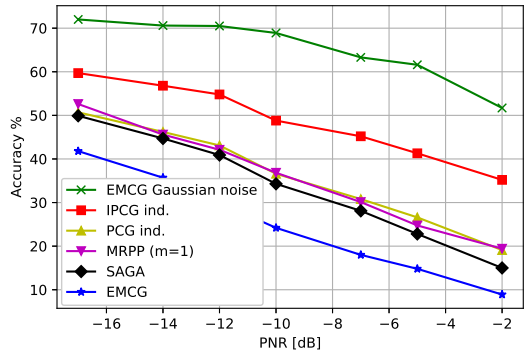}}
	\caption{Classifier accuracy under different attack schemes.}
	\label{2ant}
\end{figure}

\begin{figure}[t]
	\centerline{\includegraphics[width=0.8\linewidth]{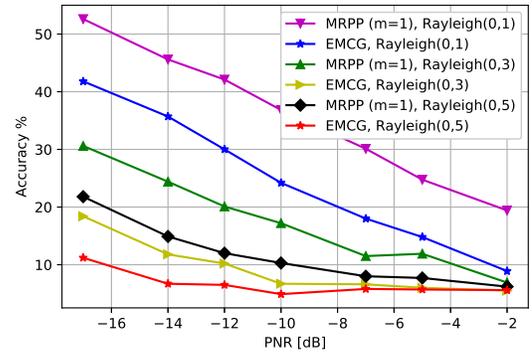}}
	\caption{Classifier accuracy with respect to different Rayleigh fading variances.}
	\label{2ant_5}
\end{figure}

First, we compare the classifier accuracy of an adversary equipped with a single antenna using the MRPP attack to the case of multiple adversaries where each adversary has a single antenna using the MRPP attack. For a fair comparison, total power that is used among adversaries is kept the same as the power used by the single adversary and the power is equally divided among adversaries. Results are shown in Fig. \ref{adv}. Note that for the case of two or more adversaries, adversaries are not synchronized and do not collaborate with each other as they are physically not co-located meaning that they attack with independent targets. We observe that the accuracy of the classifier does not drop although more adversaries are used to attack the classifier. This result suggests that dividing the power equally is not helpful and thus motivates the need for an adversary with multiple antennas to choose power allocation on antennas and exploit the channel diversity.

Adversarial attacks using two antennas with common target and independent targets are compared in Fig. \ref{target}. The PCG attack outperforms the IPCG attack regardless of whether the target is common or independent showing that the power allocation among antennas is important. Also, choosing an independent target at each antenna performs better than the common target case for both PCG and IPCG attacks suggesting that choosing the best target (determined by the channel realization) for each antenna is more effective.

\begin{figure}[t]
	\centerline{\includegraphics[width=0.8\linewidth]{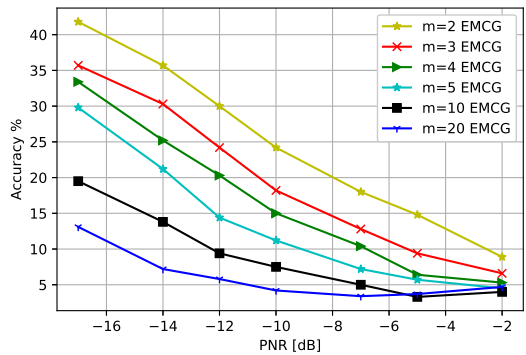}}
	\caption{Classifier accuracy with different number of antennas at the adversary.}
	\label{mul_ant}
\end{figure}
Fig. \ref{2ant} presents the classifier accuracy at the receiver when the adversary transmits an adversarial perturbation with $m=2$ antennas using different attacks that are introduced in Section \ref{sec:section3}. The EMCG attack with Gaussian noise transmitted by the adversary with two antennas is compared with the adversarial perturbation with two antennas using the MRPP attack at each antenna. The use of Gaussian noise as perturbation results in poor attack performance although the EMCG attack is used to determine the antenna to transmit supporting the use of the MRPP attack. Fig. \ref{2ant} shows that although the adversary uses two antennas, the accuracy of the classifier is higher than the case under the MRPP attack of an adversary with single antenna when the IPCG attack with independent targets is used. Also, the performance of the PCG attack with independent targets is similar to the performance of the MRPP attack of the adversary with a single antenna although the adversary puts more power to the better channel. We observe that the SAGA attack slightly outperforms the MRPP attack of an adversary with a single antenna suggesting that the SAGA attack takes advantage of having two channels to choose from. Moreover, the EMCG attack significantly outperforms other attacks by fully utilizing the channel diversity. 

So far, results have been obtained under the assumption that channels between the antennas are independent, which also yields zero covariance. Next, we consider correlation between the channels and we investigate various attacks of an adversary with two antennas under different covariance levels.  Results are shown in Fig. \ref{cor}. We observe that as the covariance between the antennas increases, the performance of the PCG attack with common target increases significantly where it is comparable to the SAGA attack and even outperforms the PCG attack with independent targets. Note that the PCG attack with independent targets outperforms the PCG attack with common target when the channels are independent as shown in Fig. \ref{target}.
 In contrast, we see that other attack schemes are not significantly affected by the covariance. Further, we observe  that even if the covariance is increased to 0.7 the attack performance slightly decreases compared to when the covariance is 0.2 in the EMCG attack, the PCG attack with independent targets and the SAGA attack.

Assuming again independent channels from adversary antennas to the receiver, the classifier accuracy is shown in Fig. \ref{2ant_5} when we vary the channel variance. The classifier accuracy drops as the channel variance increases for all cases due to the increased uncertainty induced by the increased channel gain from the adversary to the receiver. Further, the performance ratio between MRPP and EMCG attacks increases as the channel variance increases. We also observe that as the PNR increases, the gap between MRPP and EMCG attacks decreases except for the case when the channel variance is 1.


Finally, we evaluate the attack performance of the adversary with different number of antennas $m$ for the EMCG attack. Results are shown in Fig. \ref{mul_ant} when the variance of channels is 1. The classifier accuracy decreases as $m$ increases  due to the increased channel diversity available to the adversary to exploit. Moreover, as the PNR increases, the performance gap between attacks launched with different $m$ decreases suggesting that an increase of $m$ in the high PNR region is not as effective as in the low PNR region.

\section{Conclusion} \label{sec:Conclusion}
We considered a wireless communication system where a DL-based signal classifier is used at the receiver to classify signals transmitted from the transmitter to their modulation types and showed that different methods to craft adversarial perturbations can be used to exploit multiple antennas at the adversary. We show that just adding more antennas at the adversary does not always increase the attack performance. Thus, it is important to carefully allocate power among antennas, determine the adversarial perturbation for each antenna, and exploit channel diversity to select which antenna to transmit. In this context, the proposed EMCG attack significantly outperforms other attacks and effectively uses multiple antennas to evade the target classifier over the air. Next, we showed that the attack performance holds for different conditions of channels from the adversary antennas to the receiver and significantly improves by increasing the number antennas at the adversary.

%
%


\bibliographystyle{IEEEtran}
\bibliography{IEEEabrv,lib}

\end{document}